# Research and development of MolAICal for drug design via deep learning and classical programming

Qifeng Bai[1,*]

https://orcid.org/0000-0001-7296-6187

[1]School of Basic Medical Sciences, Lanzhou University, Lanzhou, Gansu 730000, P. R. China

*Corresponding author
E-mail: molaical@yeah.net

## Abstract

Deep learning methods have permeated into the research area of computer-aided drug design. The deep learning generative model and classical algorithm can be simultaneously used for three-dimensional (3D) drug design in the 3D pocket of the receptor. Here, three aspects of MolAICal are illustrated for drug design: in the first part, the MolAICal uses the genetic algorithm, Vinardo score and deep learning generative model trained by generative adversarial net (GAN) for drug design. In the second part, the deep learning generative model is trained by drug-like molecules from the drug database such as ZINC database. The MolAICal invokes the deep learning generative model and molecular docking for drug virtual screening automatically. In the third part, the useful drug tools are added for calculating the relative properties such as Pan-assay interference compounds (PAINS), Lipinski's rule of five, synthetic accessibility (SA), and so on. Besides, the structural similarity search and quantitative structure-activity relationship (QSAR), etc are also embedded for the calculations of drug properties in the MolAICal. MolAICal will constantly optimize and develop the current and new modules for drug design. The MolAICal can help the scientists, pharmacists and biologists to design the rational 3D drugs in the receptor pocket through the deep learning model and classical programming. MolAICal is free of charge for any academic and educational purposes, and it can be downloaded from the website https://molaical.github.io.

# Introduction

  With the technology development of computer-aided drug design (CADD), the deep learning method has been immersed in the research field of drug design such as the chemical synteses [1], classification of drugs [2] and so on [3]. Deep learning and classical programming of CADD have their own advantages on the development and discovery of innovative drugs. The traditional machine learning methods include the support vector machine (SVM) [4], Bayesian algorithm [5], random forest (RF) [6], artificial neural networks (ANNs) [7, 8], etc. According to some research reports, machine learning shows a good way for drug discovery. For instance, the inhibitors of kallikrein 5 protease [9] are explored via the SVM method. The binding score between ligand and protein is a key factor for screening the potential compounds. RF-based scoring functions reveal the Pearson's correlation coefficients ranging from 0.559 to 0.783 between experimental affinities and predicted values on basis of PDBbind database v2007 [10]. Some studies indicate that Bayesian algorithms can be used to identify the inhibitors of G-protein-coupled receptors [11], and so on. The ANNs have been reported to predict chemical immunotoxicity [12]. As current research reports, the deep learning method shows a better performance than the traditional machine learning methods [13-15]. The deep learning methods have popularly used in the area of drug virtual screening and *de novo* drug design [16-18]. The representative methods for *de novo* drug design include variational autoencoders (VAEs) [19] and generative adversarial net (GAN) [20, 21]. The objective-reinforced generative adversarial networks (ORGAN) is designed with the generator that produces the molecules to cheat the discriminator based on the reinforcement and adversarial learning methods [22]. The ORGAN mainly trains the deep learning model based on the SMILES sequences of ligands. The molecular graph is another way to profile the molecules with nodes and edges that correspond to the atoms and bonds of molecule. MolGAN [23] is a deep learning program that can be used to train molecular generative model based on molecular graph. The MolGAN contains GAN and VAEs for training the drug-like molecules through the annotation matrix and dense adjacency tensor. Besides, the research reports shows a good way to carry out drug virtual screening by SVM, molecular docking and molecular mechanics/generalized Born surface area (MM/GBSA), etc [24].

  The deep learning model and classical programming have different ways for drug design. The deep learning focuses on training rules of drug design based on input and output data of drugs. The well-trained model can be used to design drugs directly. In contrast, the traditional CADD software needs the algorithm rules and relative input data to design the drugs. The LigBuilder [25], OpenGrowth [26] and AutoDock Vina [27] are the popular drug design software. The LigBuilder and OpenGrowth develop the potential candidate drugs via the fragmented growth or link method that belongs to one branch of *de novo* drug design methods. The AutoDock Vina designs the ligands through molecular docking that is one way of virtual screening methods. The traditional CADD software shows a good performance for drug design. Especially, the AutoDock Vina shows the best scoring power according to the reported experiments about the ligands assessment in the receptor pocket [28]. The deep learning can learn the molecular characteristics with one-dimensional (1D) SMILES sequence, two-dimensional (2D) molecular graph and molecular graph incorporating three-dimensional (3D) structural information [29]. The classical programming can design the 3D ligands in the receptor pocket directly. In this case, the MolAICal is preliminarily designed on basis of deep learning and classical programming. The MolAICal soft package mainly focuses on three aspects of development and research. In the first part, the MolAICal is designed based on the *de novo* drug design method and deep learning model that is trained on the fragments of appointed special drugs such as Food and Drug Administration (FDA)-approved drugs. In the second part, the MolAICal contains the modules of virtual screening and deep learning model that is trained on drug-like ligands of appointed special database such as ZINC database. In the third part, the useful tools for drug filter, prediction, etc. are developed in the MolAICal software for drug discovery. The main part of MolAICal is written by JAVA program that derives many useful libraries such as Chemistry Development Kit (CDK) [30], Ambit JAVA library, etc. MolAICal uses JAVA parallelstream to invoke multicore CPU for designing drugs concurrently. MolAICal will try its best to take advantage of the drug design method of deep learning model and classical programming for drug discovery. The MolAICal supplies an effective way for 3D ligand design in the receptor pocket.

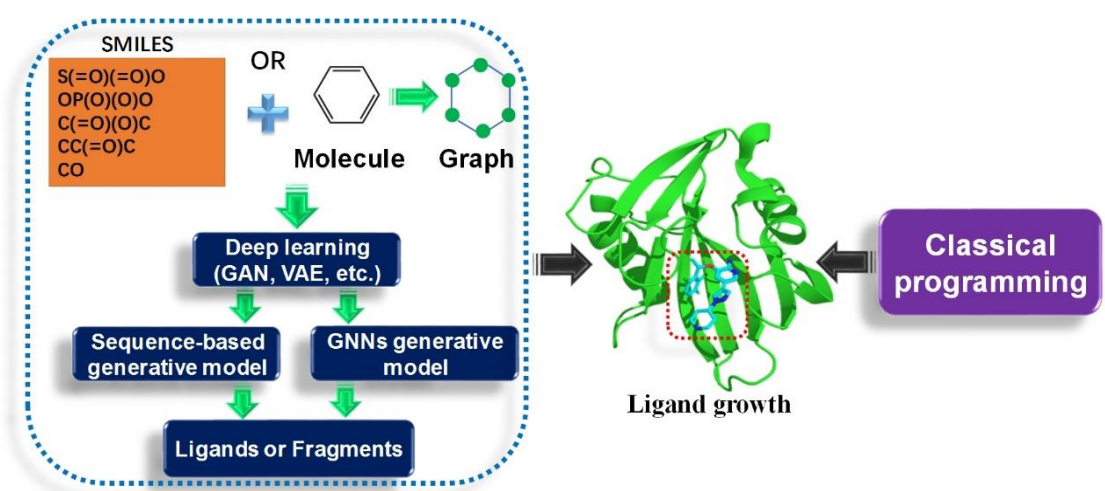

**Figure 1.** The module of MolAICal contains deep learning and classical programming methods.

## Results and Discussion

**Deep learning and *de novo* methods for drug design**

In the first module, MolAICal focuses on the method development of deep learning generative model and *de novo* drug design. As shown in Figure 1, the deep learning generative model is an important direction of research and development of MolAICal based on SMILES sequence and molecular graph. The effective and stable training methods can improve the accuracy rate of drug discovery. The popular training methods such as GAN are picked up for training the generative model of ligands and fragments in the current version of MolAICal. The MolAICal will further optimize the current training method GAN, and employ the new advanced technology for training generative models. In addition, the spatial structural information of ligand and receptor is an important factor for rational drug design. The deep learning generative model incorporating 3D structural information can show the state-of-the-art way to produce the molecules with high 3D similarity to the determined ligand of the receptor. MolAICal will further develop and train the excellent generative model incorporating 3D structural information for drug design. Meanwhile, the

MolAICal also employs classical programming such as fragment growth for drug design based on the output fragments of deep learning generative model (see Figure 1). In the process of the drug design of MolAICal, the initial fragment is chosen as the started fragment in the 3D pocket of the receptor. The next fragment grows on the previous ligand fragment. The growth algorithm and binding score play an important role in the evaluation of drug design. MolAICal will further optimize the growth algorithm and develop the superior binding score based on the PDBbind database. Besides, MolAICal also supplies the user-defined seed fragments for *de novo* drug design. Generally, deep learning and classical programming are the development aspects of the MolAICal. The ultimate goal is that MolAICal can try its best to take advantage of the deep learning and classical programming for drug design.

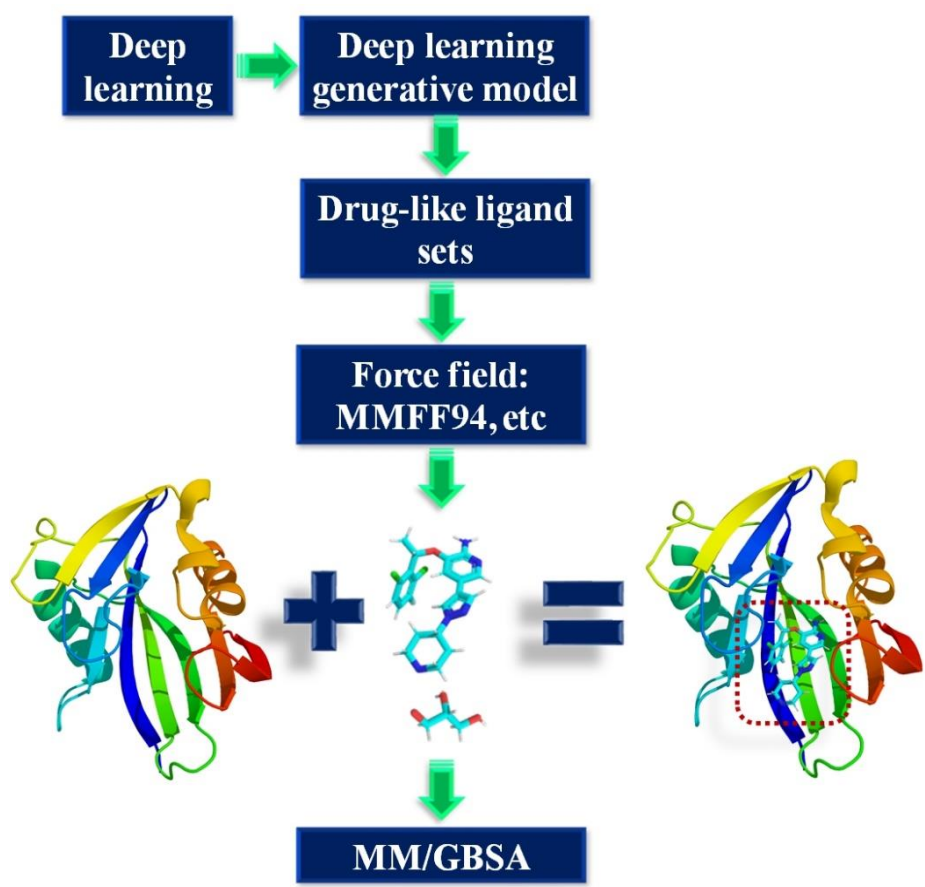

**Figure 2.** The module of MolAICal contains deep learning and drug virtual screening method.

**Deep learning and virtual screening method for drug design**

In the second module, the deep learning generative model and virtual screening method are introduced for drug screening in the MolAICal. MolAICal provides two drug design ways that screen ligands in the 3D pocket of receptors based on the known drug database and generated database by deep learning model. First of all, the high-quality ligand sets can promote the speed of new drug discovery. As shown in Figure 2, MolAICal employs the deep learning generative model for producing the drug-like database. The deep learning generative model can generate high-quality, novel, synthesizable, diversity and drug-like ligands. Secondly, because the ligands are the SMILES format in the generated drug database, the software Open Babel [31] is employed to transform the

3D structure from the SMILES format ligands on basis of Merck Molecular Force Field 94 (MMFF94). The MolAICal will further develop its module to satisfy the requirement of molecular format conversion for drug design. Thirdly, the drug virtual screening is performed on the 3D receptor pocket based on the generated drug database by invoking the molecular docking program Autodock Vina (see Figure 2). The MolAICal will further update Autodock Vina to another version named MACDock. The MACDock will improve its performance based on "scoring power", "ranking power", "docking power", and "screening power". In addition, drug virtual screening based on the known ligands database is also added in MolAICal. Lastly, the MM/GBSA is a kind of recognized accurate way to assess the binding free energy of protein-ligand, protein-protein, and so on [32]. MolAICal provides a way to calculate the MM/GBSA based on the output results of molecular dynamics simulations that are carried out by software NAMD. The score functions which can be used to assess the affinities of ligands in the receptor pocket are an important development area in MolAICal. The MolAICal will develop its molecular format transformation and 3D coordinate generation based on the force field. The drug generative model and molecular docking program, etc are also added in the development plan of MolAICal.

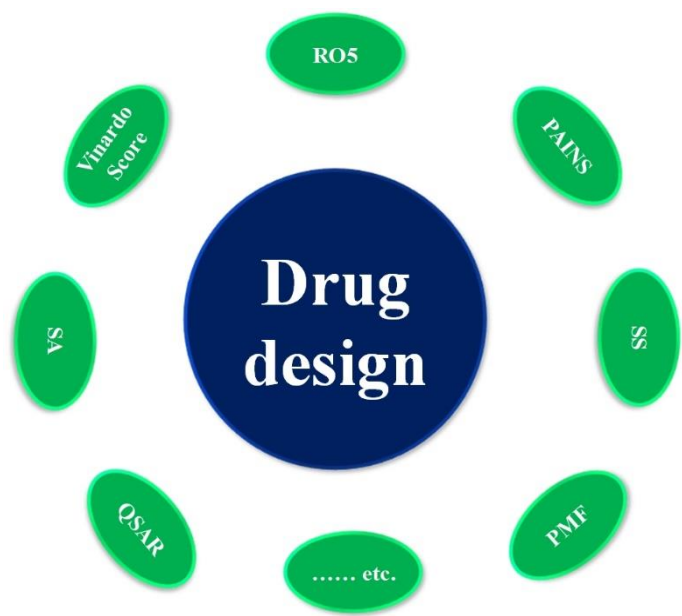

**Figure 3.** The useful tools of MolAICal for drug design.

**Useful tools for drug design**

The deep learning model, classical programming and Vinardo score can be used to select the good affinity ligands in soft package MolAICal. However, it still needs to evaluate absorption, distribution, metabolism, excretion, toxicity, drug-likeness, etc of drugs. In the current version of MolAICal, Lipinski's rule of five (RO5), pan-assay interference compounds (PAINS) and synthetic accessibility (SA) are embedded to assess the drug-likeness of ligands (see Figure 3). The parameters and methods of these filter rules will further be optimized and improved in MolAICal. For example, as the research report of FDA approved oral drugs [33], the RO5 has substantially changed its cutoff values of rotatable bonds, molecular weight, and hydrogen bond acceptors. In this case, the parameters of RO5 need to be updated according to the research report. Besides, MolAICal also contains useful drug design tools such as quantitative structure-activity relationship (QSAR), ligand similarity search (SS) and potential of mean force (PMF) (see Figure 3). These kinds of tools can be quickly used to screen or evaluate the drugs. The MolAICal will further develop the medicinal chemistry filters (MCFs) [34], descriptor MCE-18 [35], weak interaction [36] and so on. Generally, MolAICal will continually develop new and effective tools for drug filter and design.

## Conclusions

In this paper, the main three modules of MolAICal are introduced for drug design. For the first module, the deep learning and *de novo* methods are developed for fragment growth in the 3D pocket of the receptor. For the second module, deep learning and virtual screening method are added for drug screening in the 3D pocket of the receptor. Both of two modules in MolAICal can be also used to design drugs in the receptor pocket based classical programming independently. For the third module, the MolAICal focuses on the development of effective and useful tools for drug filter and design. In total, the MolAICal will continually optimize and develop new methods for drug design. The ultimate aim is that MolAICal can help scientists to find new drugs efficiently.

## Materials and methods

### Deep learning

In the current version, the Wasserstein generative adversarial networks (WGANs) [37] are picked up to train the deep learning generative model of MolAICal. The WGANs can be used to calculate the difference of probability distributions between the fraud and genuine drugs data distribution (see equation 1).

$$W(\mathbb{P}_r, \mathbb{P}_\theta) = \sup_{||f||_L \leq 1} \mathbb{E}_{x \sim \mathbb{P}_r}[f(x)] - \mathbb{E}_{x \sim \mathbb{P}_\theta}[f(x)] \qquad (1)$$

To obtain the best optimization of equation 1, the solution of WGANs is shown in equation 2:

$$\min_G \max_D \mathbb{E}_{x \sim \mathbb{P}_r}[D(x)] - \mathbb{E}_{z \sim p(z)}[D(G(z))] \qquad (2)$$

Where the *D* and *G* are the discriminator and generator, respectively. The discriminator *D* increases the distinguishable probability of genuine data and decrease the indistinguishable probability of fraud data in the training process. The stable training WGANs is added to the development plan of MolAICal.

### Classical programming

The genetic algorithm is used for *de novo* drug design. The initial fragment is chosen from the part of known ligand in the receptor pocket or generated by appointed SMILES around the atom of key residue. The next fragment grows on the previous fragment through the perturbation search of the Fibonacci or random algorithm. The Fibonacci points are distributed on the sphere according to equation 3:

$$x_i = \sqrt{1 - \left(1 - \frac{1}{N} - \frac{2i}{N}\right)^2} * \cos(\pi(3 - \sqrt{5}) * i)$$

$$y_i = \sqrt{1-\left(1-\frac{1}{N}-\frac{2i}{N}\right)^2} * \sin(\pi(3-\sqrt{5})*i) \qquad (3)$$

$$z_i = 1 - \frac{1}{N} - \frac{2i}{N}$$

Where $x_i$, $y_i$ and $z_i$ represent the position of sphere points in the coordinate system. The letter $N$ represents the number of produced points that can be appointed by users. The operators' crossover and mutation of genetic algorithm (GA) are used to optimize the ligands in the receptor pocket. The MolAICal also employs classical programming for drug virtual screening and unwanted ligands filtering. MolAICal will optimize these methods, and further add the new effective classical programming for drug design.